\documentclass[dvips]{acta}
\usepackage{supertabular,lscape,epsfig}
\usepackage{amssymb}
\usepackage{amsmath}
\DeclareSymbolFont{ppa}{OT1}{ppl}{m}{it}
\DeclareMathSymbol{\vv}{\mathalpha}{ppa}{'166}

\newfont{\hb}{rphvb at 10pt}
\newfont{\hbo}{rphvbo at 10pt}
\newfont{\bitt}{rptmbi at 12pt}
\newfont{\bits}{rptmbi at 11pt}

\SetPages{1}{1}

\SetVol{65}{2015}

\begin{document}

\newcommand{\TabCapp}[2]{\begin{center}\parbox[t]{#1}{\centerline{
  \small {\spaceskip 2pt plus 1pt minus 1pt T a b l e}
  \refstepcounter{table}\thetable}
  \vskip2mm
  \centerline{\footnotesize #2}}
  \vskip3mm
\end{center}}

\newcommand{\TTabCap}[3]{\begin{center}\parbox[t]{#1}{\centerline{
  \small {\spaceskip 2pt plus 1pt minus 1pt T a b l e}
  \refstepcounter{table}\thetable}
  \vskip2mm
  \centerline{\footnotesize #2}
  \centerline{\footnotesize #3}}
  \vskip1mm
\end{center}}

\newcommand{\MakeTableSepp}[4]{\begin{table}[p]\TabCapp{#2}{#3}
  \begin{center} \TableFont \begin{tabular}{#1} #4
  \end{tabular}\end{center}\end{table}}

\newcommand{\MakeTableee}[4]{\begin{table}[htb]\TabCapp{#2}{#3}
  \begin{center} \TableFont \begin{tabular}{#1} #4
  \end{tabular}\end{center}\end{table}}

\newcommand{\MakeTablee}[5]{\begin{table}[htb]\TTabCap{#2}{#3}{#4}
  \begin{center} \TableFont \begin{tabular}{#1} #5
  \end{tabular}\end{center}\end{table}}

\newfont{\bb}{ptmbi8t at 12pt}
\newfont{\bbb}{cmbxti10}
\newfont{\bbbb}{cmbxti10 at 9pt}
\newcommand{\uprule}{\rule{0pt}{2.5ex}}
\newcommand{\douprule}{\rule[-2ex]{0pt}{4.5ex}}
\newcommand{\dorule}{\rule[-2ex]{0pt}{2ex}}
\def\thefootnote{\fnsymbol{footnote}}
\begin{Titlepage}
\Title{The OGLE Collection of Variable Stars.\\
Anomalous Cepheids in the Magellanic Clouds\footnote{Based on observations obtained with the 1.3-m Warsaw
telescope at the Las Campanas Observatory of the Carnegie Institution
for Science.}}
\Author{I.~~S~o~s~z~y~ñ~s~k~i$^1$,~~
A.~~U~d~a~l~s~k~i$^1$,~~
M.\,K.~~S~z~y~m~a~ñ~s~k~i$^1$,~~
G.~~P~i~e~t~r~z~y~ñ~s~k~i$^1$,\\
\L.~~W~y~r~z~y~k~o~w~s~k~i$^1$,~~
K.~~U~l~a~c~z~y~k$^2$,~~
R.~~P~o~l~e~s~k~i$^{1,3}$,~~
P.~~P~i~e~t~r~u~k~o~w~i~c~z$^1$,\\
S.~~K~o~z~³~o~w~s~k~i$^1$,~~
J.~~S~k~o~w~r~o~n$^1$,~~
D.~~S~k~o~w~r~o~n$^1$,~~
P.~~M~r~ó~z$^1$,~~
and~~M.~~P~a~w~l~a~k$^1$}
{$^1$Warsaw University Observatory, Al.~Ujazdowskie~4, 00-478~Warszawa, Poland\\
e-mail: (soszynsk,udalski)@astrouw.edu.pl\\
$^2$ Department of Physics, University of Warwick, Gibbet Hill Road, Coventry, CV4 7AL, UK\\
$^3$ Department of Astronomy, Ohio State University, 140 W. 18th Ave., Columbus, OH 43210, USA}
\Received{~}
\end{Titlepage}
\Abstract{We present a collection of 250 anomalous Cepheids (ACs)
discovered in the OGLE-IV fields toward the Large (LMC) and Small
Magellanic Cloud (SMC). The LMC sample is an extension of the OGLE-III
Catalog of ACs published in 2008, while the SMC sample contains the first
known bona fide ACs in this galaxy. The total sample is composed of 141 ACs
in the LMC and 109 ACs in the SMC. All these stars pulsate in single modes:
fundamental (174 objects) or first overtone (76 objects). Additionally, we
report the discovery of four ACs located in the foreground of the
Magellanic Clouds. These are the first fundamental-mode ACs known in the
Galactic field.

We demonstrate that the coefficients $\phi_{21}$ and $\phi_{31}$ determined
by the Fourier light curve decomposition are useful discriminators between
classical Cepheids and ACs, at least in the LMC and in the field of the
Milky Way. In the SMC, the light curve shapes and mean magnitudes of
short-period classical Cepheids make them similar to ACs, which is a source
of difficulties in the discrimination of both classes of pulsators. The
presence of unidentified ACs in the catalogs of classical Cepheids may be
partly responsible for the observed non-linearity of the period-luminosity
relation observed for short-period Cepheids in the SMC. We compare spatial
distributions of ACs, classical Cepheids and RR Lyr stars. We show that the
distribution of ACs resembles that of old stars (RR Lyr variables),
although in the LMC there are visible structures typical for young
population (classical Cepheids): the bar and spiral arms. This may suggest
that ACs are a mixture of relatively young stars and mergers of very old
stars.} {stars: variables: Cepheids -- Stars: oscillations -- Stars:
Population II -- Magellanic Clouds}

\Section{Introduction}

Although anomalous Cepheids (ACs) have been observed and studied for over
half a century, their origin is still a subject of debate. There is a
consensus that ACs are metal-deficient core-helium-burning pulsating stars
with masses 1-2~$M_{\odot}$ (\eg Bono \etal 1997, Caputo \etal 2004,
Marconi \etal 2004). They occur in all dwarf galaxies that have been
searched for variable stars and are remarkably rare in globular
clusters. First representatives of ACs were discovered by Thackeray (1950)
in the Sculptor dwarf and by Baade and Swope (1961) in the Draco dwarf
galaxy. The term ``anomalous Cepheids'' was introduced by Zinn and Searle
(1976), since the properties of these variables do not match neither those
of classical nor type II Cepheids.

First unambiguously detected ACs in the Large Magellanic Cloud (LMC) were
reported by Soszyñski \etal (2008) on the basis of the photometric
observations obtained during the third phase of the Optical Gravitational
Lensing Experiment (OGLE-III). The OGLE sample consisted of 83 variables in
the LMC (62 funda\-mental-mode and 21 first-overtone pulsators), which
nearly doubled the total number of all ACs known at that time in the
Universe. This sample was used in several important researches.

Fiorentino and Monelli (2012) estimated that masses of ACs in the LMC range
from 0.8 to 1.8~$M_{\odot}$, with a mean value of $1.2\pm0.2
M_{\odot}$. They also compared the spatial distribution of ACs, classical
Cepheids, type II Cepheids and RR~Lyr stars, all originated from the
OGLE-III Catalog of Variable Stars in the LMC. Their investigation could
potentially answer the question how ACs were formed: are they
intermediate-age stars (1-6 Gyr old) with exceptionally low metallicity
(Demarque and Hirshfeld 1975, Norris and Zinn 1975) or are they coalesced
old (>10 Gyr) binary stars (Renzini \etal 1977)? Fiorentino and Monelli
(2012) found that the distribution of ACs is different from both: young
stellar population represented by classical Cepheids and old stars
indicated by RR Lyr stars and type II Cepheids. They concluded that a
survey of the outskirts of the LMC would probably solve the problem of the
ACs origin.

Recently, Ripepi \etal (2014) cross-correlated the OGLE catalog of ACs
against the near-infrared $K_s$-band light curves collected by the VMC
survey (Cioni \etal 2011) and derived period--luminosity (PL) relations for
these pulsators. First application of the PL relations to ACs known in
nearby dwarf galaxies revealed that they could be important distance
indicators within the Local Group. OGLE ACs in the LMC provide currently
the best pattern for the PL relations of this type of variables (\eg Osborn
\etal 2012, Sipahi \etal 2013a, Cusano \etal 2013).

ACs in the LMC form a well separated group of pulsating stars, with PL
relations located between classical and type II Cepheids. In contrast, we
did not find such a well-defined group of ACs in the Small Magellanic Cloud
(SMC). Soszyñski \etal (2010) listed only six candidates for ACs (3
fundamental-mode and 3 first-overtone pulsators) in the SMC. In this paper,
we report the discovery of the first bona fide ACs in this galaxy. We found
that the SMC ACs are very similar to classical Cepheids in the sense of
their mean brightness and light curve morphology. We also extend the
OGLE-III sample of ACs in the LMC by new identifications in the outer
regions of this galaxy observed by the fourth phase of the OGLE project
(OGLE-IV, Udalski \etal 2015).

\Section{Observational Data}

Time-series {\it I} and {\it V}-band photometry of the Magellanic Clouds
was obtained in the years 2010--2015 using the 32-chip mosaic CCD camera
mounted at the focus of the Warsaw Telescope located at Las Campanas
Observatory in Chile. The observatory is operated by the Carnegie
Institution for Science. The OGLE-IV camera has a total field of view of
1.4~square degrees and pixel scale of 0.26~arcsec. The OGLE-IV fields cover
approximately 650~square degrees in both Clouds and a region between both
galaxies, the so-called Magellanic Bridge. For each field we obtained from
90 (in sparse regions far from the centers of the Magellanic Clouds) to
over 750 observing points (in the densest fields) in the Cousins {\it I}
band and from several to over 260 points in the Johnson {\it V} band.

Data reduction of the OGLE images was performed using the Difference Image
Analysis technique (Alard and Lupton 1998, Wo¼niak 2000). Detailed
descriptions of the instrumentation, photometric reductions and astrometric
calibrations of the OGLE-IV data are provided by Udalski \etal (2015).

\Section{Identification and Classification of Anomalous Cepheids}

The first step in the identification of anomalous Cepheids was to search
for periods for all {\it I}-band light curves of stellar objects collected
in the Magellanic System by the OGLE-IV survey. The frequency analysis for
75 million stars was performed using the {\sc Fnpeaks}
program\footnote{http://helas.astro.uni.wroc.pl/deliverables.php?lang=en\&active=fnpeaks}
kindly provided by Z.~Ko³aczkowski. We tested frequencies from 0 to
24~d$^{-1}$ with a spacing of 0.00005~d$^{-1}$. Light curves with the
largest signal to noise ratios of the detected periods and those which were
located in the ``Cepheid region'' in the PL diagram (wide strip covering
classical and type II Cepheids) were subjected to visual inspection.

\begin{figure}[p]
\includegraphics[width=12.7cm]{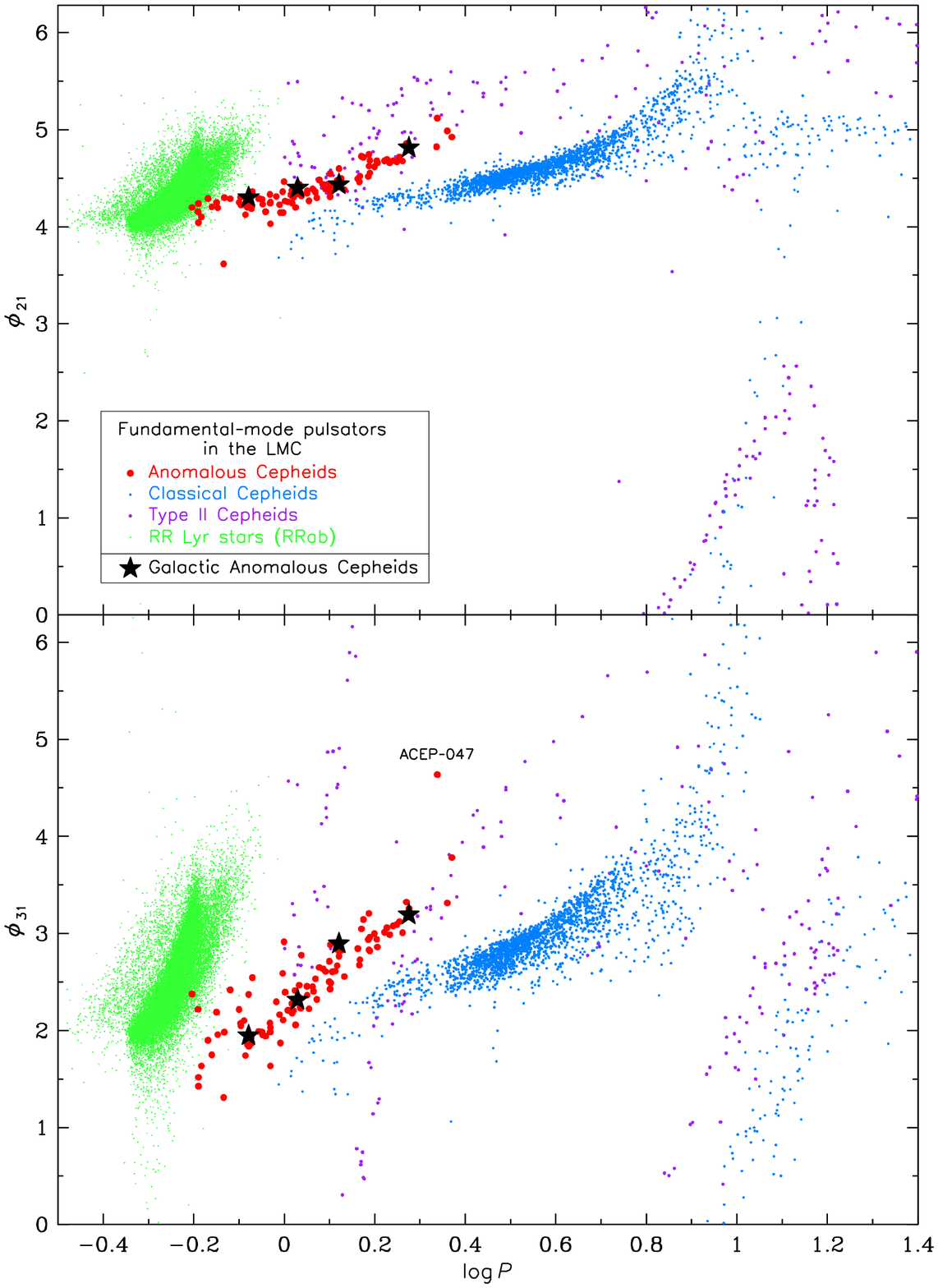}
\FigCap{Fourier coefficients $\phi_{21}$ and $\phi_{31}$ as a function of
periods for the fundamental-mode Cepheids and RR Lyr stars detected in the
OGLE fields toward the LMC. Red, blue, purple and green points indicate
ACs, classical Cepheids, type II Cepheids, and RR Lyr stars,
respectively. Four black star symbols show Galactic ACs found in the
foreground of the Magellanic Clouds.}
\end{figure}

\begin{figure}[p]
\includegraphics[width=12.7cm]{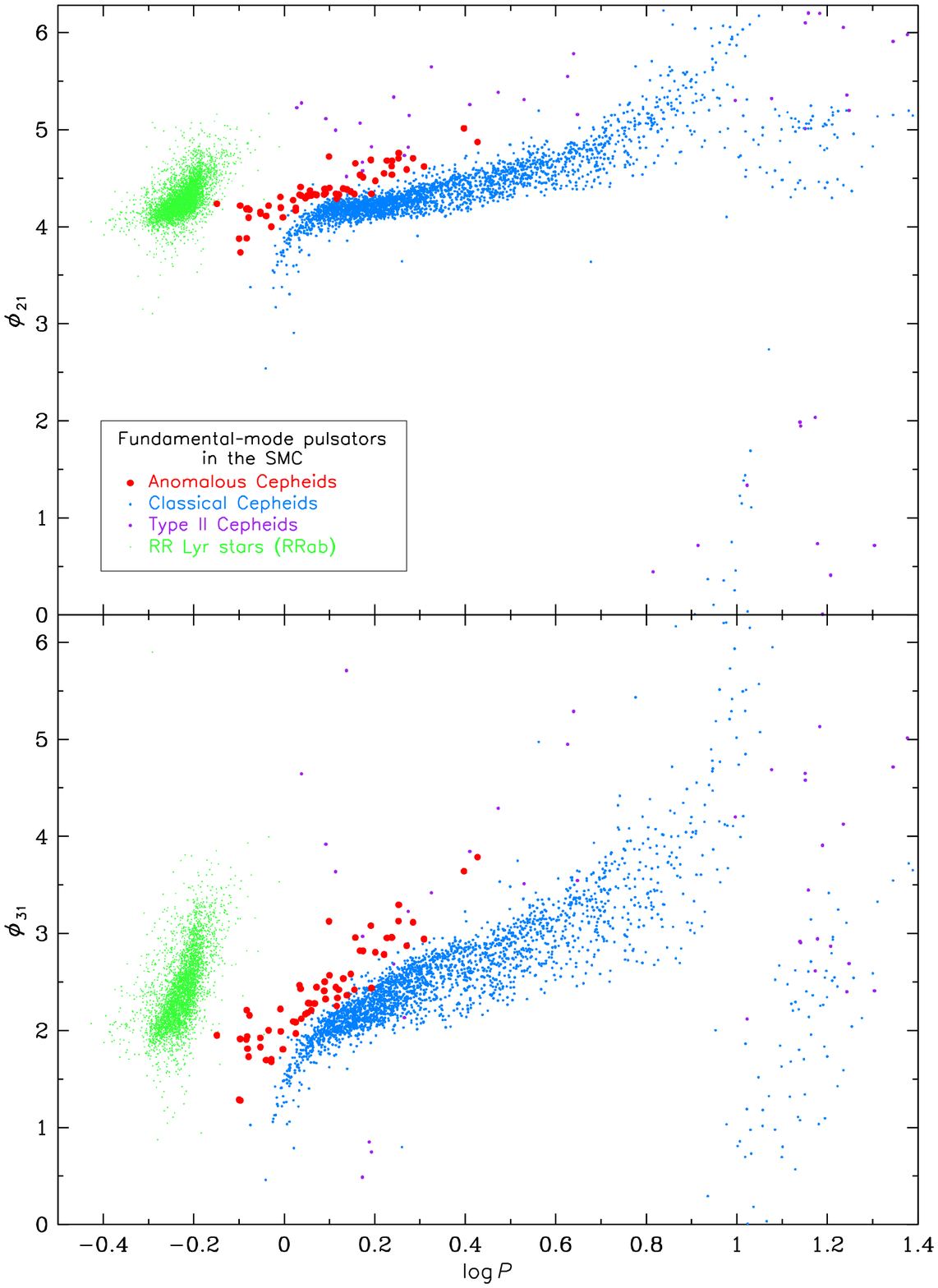}
\FigCap{Fourier coefficients $\phi_{21}$ and $\phi_{31}$ as a function of
periods for the fundamental-mode Cepheids and RR Lyr stars detected in the
OGLE fields toward the SMC. Different colors of points represent the same
types of variable stars as in Fig.~1.}
\end{figure}

Variable stars were initially divided into three groups: pulsating,
eclipsing, and other (usually of unknown type) variables. A more detailed
classification of the first group was performed on the basis of the
position of stars in the PL diagram and Fourier parameters fitted to the
light curves. In general, ACs can be confused with RR Lyr stars,
classical and type II Cepheids, although in the LMC they form a quite well
separated group. Soszyñski \etal (2008) demonstrated that ACs in the LMC
delineate PL relations located between classical Cepheids and type II
Cepheids/RR Lyr stars. In Fig.~1 we show that fundamental-mode ACs in the
LMC are characterized also by a distinct light curve morphology. Here
$\phi_{21}$ and $\phi_{31}$ are phase differences of the Fourier cosine
series fitted to the {\it I}-band light curves. It is clear that ACs (red
points) can be well distinguished from classical Cepheids (blue points)
only on the basis of their light curves.

The situation looks different in the SMC. Soszyñski \etal (2010) noticed
that there is no distinct group of pulsating stars between classical and
type II Cepheids in the PL plane. However, it is known that
fundamental-mode classical Cepheids in the SMC exhibit a break in the slope
of the PL relation for a period of about 2.5~days (Bauer \etal 1999,
Udalski \etal 1999), which may be also interpreted as an excess of faint
Cepheids at the short-period end of the PL relation. We found that this
excess is even more pronounced far from the SMC center, in the regions not
observed by the OGLE-III project, but regularly monitored by the OGLE-IV
survey.

A closer look at these faint Cepheids revealed that most of them have
different light curves than regular classical Cepheids, although this
difference is not so obvious as in the case of ACs and classical Cepheids
in the LMC. Fig.~2 is analogous to Fig.~1, but presents variables in the
SMC. Fundamental-mode Cepheids located distinctly below the mean PL
relations of classical Cepheids and with $\phi_{21}$ and $\phi_{31}$
parameters larger than typical values for classical Cepheids were
classified as ACs. A similar procedure was performed for first-overtone
Cepheids, although their light curves exhibit larger diversity then their
fundamental-mode counterparts, so our classification is more
uncertain. Since ACs and classical Cepheids partly overlap in both -- PL
and period \vs Fourier coefficients -- diagrams, our classification may be
incorrect for individual stars, both fundamental-mode and overtone
pulsators. On the other hand, there is no doubt that ACs, although
morphologically similar to classical Cepheids, exist in the SMC, which is
further confirmed by their radically different spatial distribution (see
Section 6.2).

\Section{Catalog of Anomalous Cepheids in the Magellanic Clouds}

The total sample of ACs in the Magellanic Clouds, including those detected
in the OGLE-III survey (Soszyñski \etal 2008), consists of 250 objects. The
exact number of ACs belonging to the LMC and SMC cannot be given, because
several pulsators were found in the Magellanic Bridge, approximately
halfway between both galaxies. We allocated these Magellanic Bridge ACs to
the LMC (2 objects) or SMC (14 objects), according to their position in the
sky and the position in the PL diagrams (which indicates their distance
from us), but our assignments should be treated with caution. Thus, we
identified 141 ACs in the LMC (101 fundamental-mode and 40 first-overtone
pulsators) and 109 ACs in the SMC (73 fundamental-mode and 36
first-overtone). No double-mode ACs have been found. We also detected four
pulsating stars with light curves typical for fundamental-mode ACs, but
much brighter than their counterparts in the Magellanic Clouds. These are
likely Galactic ACs located in the foreground of these galaxies
(Section~6.3).

The entire collection of ACs in the Magellanic Clouds can be downloaded from
the OGLE Internet Archive:
\begin{center}
{\it ftp://ftp.astrouw.edu.pl/ogle/ogle4/OCVS/lmc/acep/}\\
{\it ftp://ftp.astrouw.edu.pl/ogle/ogle4/OCVS/smc/acep/}
\end{center}
for the LMC and SMC samples, respectively. Each AC has its unique
identifier of the form OGLE-LMC-ACEP-NNN or OGLE-SMC-ACEP-NNN (where NNN is
a three-digit sequential number), compatible with the identifiers used in
the OGLE-III catalog (Soszyñski \etal 2008).

In the LMC, ACs numbered from OGLE-LMC-ACEP-001 to OGLE-LMC-ACEP-083 were
discovered and cataloged by Soszyñski \etal (2008). In the current version
of the catalog we removed two objects: OGLE-LMC-ACEP-022 and
OGLE-LMC-ACEP-083, since we reclassified them as RR Lyr stars. On the other
hand, a star previously classified as a type II Cepheid
(OGLE-LMC-T2CEP-114) were moved to the collection of ACs in the LMC. Newly
detected ACs in the LMC have identifiers between OGLE-LMC-ACEP-084 and
OGLE-LMC-ACEP-143 and are organized by increasing right ascension.

\begin{figure}[t]
\includegraphics[width=12.7cm]{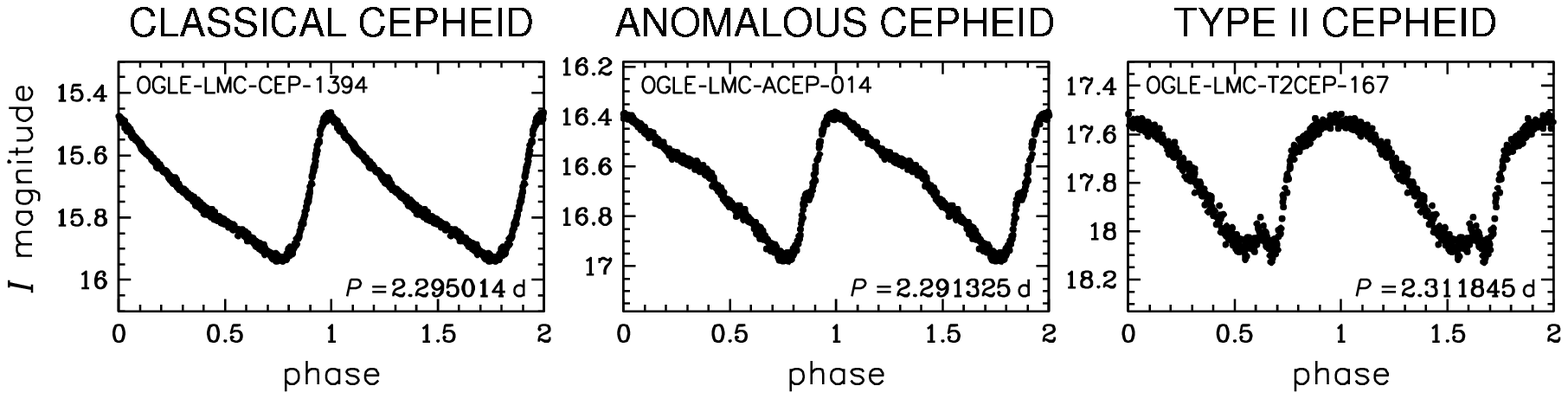}
\FigCap{Typical {\it I}-band light curves of classical Cepheids, ACs, and
type II Cepheids (BL Her stars) with periods of about 2.3~d in the LMC.}
\end{figure}

Ripepi \etal (2014) suggested that three variables with periods longer than
2~d (OGLE-LMC-ACEP-014, 033, and 047), classified by Soszyñski \etal (2008)
as ACs in the LMC, are rather BL Her stars (type II Cepheids). We carefully
inspected their light curves and we maintain our classification for
OGLE-LMC-ACEP-014 and 033. The light curves of BL Her stars with periods
above 2~d have completely different morphology (compare typical light
curves of classical, anomalous and type II Cepheids with periods of about
2.3~d in Fig.~3). Also the mean luminosity of these two ACs, although
slightly lower than suggested by the linear fit to the PL relation for
fundamental-mode ACs, is by at least 0.7~mag higher than expected for type
II Cepheids of the same periods.

\begin{figure}[t]
\begin{center}
\includegraphics[width=10cm]{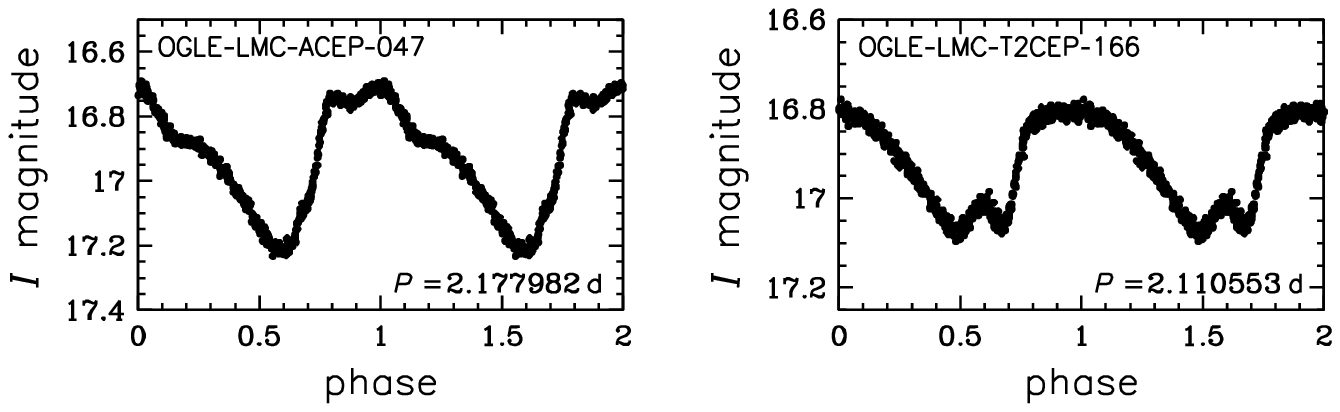}
\end{center}
\vspace*{-3mm}
\FigCap{{\it Left panel}: {\it I}-band light curve of OGLE-LMC-ACEP-047
-- a Cepheid intermediate between ACs and BL Her stars. {\it Right panel}:
{\it I}-band light curve of a blended BL Her star OGLE-LMC-T2CEP-166.}
\end{figure}

The third of the Cepheids reclassified by Ripepi \etal (2014) --
OGLE-LMC-ACEP-047 -- is an exceptional case of a pulsating star
intermediate between anomalous and type II Cepheids. In the PL diagram it
is located halfway between the relations of both types of Cepheids. Its
light curve (left panel of Fig.~4) also differs from those of ACs or BL~Her
stars (Fig.~3), since OGLE-LMC-ACEP-047 has a distinct bump on the
descending branch near the light maximum. This feature affects the
$\phi_{31}$ Fourier coefficient, which shift this star outside the region
occupied by typical ACs and BL Her stars (Fig.~1). Thus, we do not confirm
that OGLE-LMC-ACEP-047 is a typical BL~Her star and we temporarily leave
it on the list of ACs, although we point out that this star deserves
special attention in the future.

In turn, Fiorentino and Monelli (2012) reclassified one of the OGLE type II
Cepheids in the LMC (namely OGLE-LMC-T2CEP-166) as an AC. Indeed, this star
is brighter than other type II Cepheids with the same periods and falls
among ACs in the PL diagram, but the light curve of OGLE-LMC-T2CEP-166
(right panel of Fig.~4) undoubtedly indicates that this star is a type II
Cepheid. Its higher luminosity and decreased amplitude of light variations
(if expressed in magnitudes) is most likely associated with the blending by
an unresolved star.

In the SMC, we did not hitherto published a separate catalog of ACs,
although six candidates for ACs were listed by Soszyñski \etal (2010). All
these objects are included in the present collection. Our final sample
contains a total of 41 pulsating stars from the SMC classified by Soszyñski
\etal (2010) as classical Cepheids and now reclassified as ACs. We provide
their previous designations in the remarks file of the catalog.

In the FTP site we provide OGLE-IV multi-epoch photometry in the {\it I}
and {\it V} filters, finding charts, and basic observational parameters of
each AC: coordinates, periods, intensity mean magnitudes in the {\it I}-
and {\it V}-bands, {\it I}-band peak-to-peak amplitudes, and Fourier
coefficients $R_{21}$, $\phi_{21}$, $R_{31}$, and $\phi_{31}$ derived for
the {\it I}-band light curves. The pulsation periods were refined with the
{\sc Tatry} code (Schwarzenberg-Czerny 1996) using OGLE-IV observations
obtained between 2010 and 2014. Note that ACs in the central regions of the
LMC and SMC were also observed during the OGLE-II (1997-2000) and/or
OGLE-III (2001-2009) surveys, and these light curves were published by
Soszyñski \etal (2008, 2010). We suggest to merge both datasets, if one
plans to study long-term behavior of ACs in the Magellanic Clouds. Several
ACs discovered by Soszyñski \etal (2008) do not have OGLE-IV observations,
because they fell in the gaps between the CCD chips of the mosaic
camera. For these stars we provide their parameters from the OGLE-III
catalog.

\Section{Completeness of the Sample}

Some stars in our sample have two entries in the OGLE-IV database, because
they fell in the overlapping regions of two neighboring fields. In such
cases, we publish only one light curve, usually that with the larger number
of data points. These double detections can be used to test the
completeness of our collection of ACs in the Magellanic Clouds. We {\it a
posteriori} checked that 27 of the 250 ACs should be detected twice, in the
overlapping parts of the adjacent fields, so we had an opportunity to
identify 54 counterparts. We independently detected 53 of them. The only
missing light curve consists of only 20 observing points and was not
searched for periodicity in the first step of our procedure. It seems that
the completeness of our sample is very high, close to 100\%.

One should be aware that the very high level of completeness refers to the
regions covered by the OGLE fields. The sky coverage outside the central
regions of both Magellanic Clouds has many gaps resulting from the
observing strategy. Moreover, in the OGLE-IV mosaic camera there are
technical gaps between CCD detectors, which decrease an effective area of
the OGLE-IV fields by about 6\%. Of course, these restrictions do not
affect central regions of the Clouds, which were observed during the
OGLE-II and OGLE-III projects. In these areas we expect the largest level of
completeness, since it was searched for variable stars in the past (\eg
Soszyñski \etal 2008, 2010).

We cross-matched our collection of ACs with the extragalactic part of the
General Catalogue of Variable Stars (GCVS, Artyukhina \etal 1995). We found
nine counterparts, usually classified as RR Lyr stars or classical
Cepheids. Only one object -- CV101 = OGLE-SMC-ACEP-048 discovered by Landi
Dessy (1959) -- is classified in the GCVS as a BL Boo star (an alternative
name for ACs). It is interesting that GCVS lists as many as 41 stars in the
SMC classified as BL Boo stars. These objects were discovered half a
century ago by Landi Dessy (1959), Tifft (1963), Wesselink and Shuttleworth
(1965), and Graham (1975). All but one of these stars were observed by OGLE
and only CV101 is a real AC. From the remaining 39 stars designated as
BLBOO in the GCVS over 75\% turned out to be classical Cepheids in the
SMC. In many cases the pulsation periods provided in the GCVS are daily
aliases of the real periods. Several of the misclassified variables are
Galactic RR Lyrae stars in the foreground of the SMC, one star (HV~12132)
is an eclipsing binary system, and one object (HV~12912) seems to be a
constant star in the OGLE database. HV12153 suggested by Duncan \etal
(1993) to be an AC turned out to be a typical first-overtone classical
Cepheid.

\Section{Discussion}

\Subsection{Period--Luminosity Relations}

ACs, as other types of pulsating stars, obey PL laws. The OGLE collection
of ACs in the Magellanic Cloud offers an opportunity to study in detail
these relations, since both galaxies host rich samples of various pulsating
stars and other distance indicators. Baade and Swope (1961) first suggested
that ACs may obey a PL law different than classical and type II
Cepheids. This hypothesis was confirmed by van Agt (1967) and Swope
(1968). Then, the PL relations for ACs were studied, among others, by
Pritzl \etal (2002), Marconi \etal (2004), and Ripepi \etal (2014).

\begin{figure}[p]
\includegraphics[width=12.7cm]{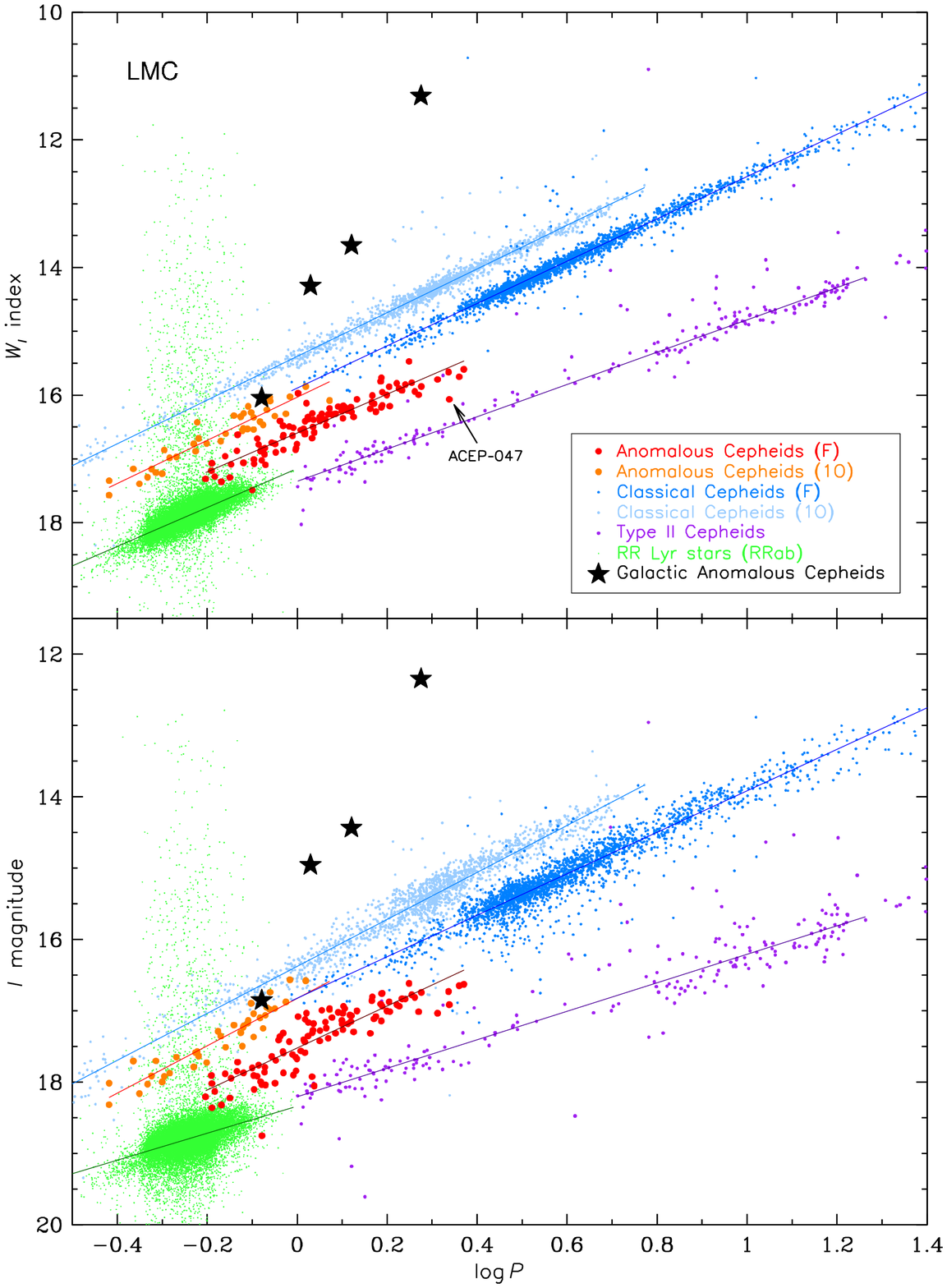}
\FigCap{Period--luminosity diagrams for ACs (red and orange points),
classical Cepheids (blue points), type II Cepheids (purple points) and
RR~Lyr stars (green points) in the LMC. Four black star symbols show
Galactic ACs in the foreground of the Magellanic Clouds. {\it Upper panel}
shows the $\log{P}$--$W_I$ relations, where $W_I=I-1.55(V-I)$ is an
extinction-free Wesenheit index. {\it Lower panel} shows the $\log{P}$--$I$
relations, where $I$ are apparent mean magnitudes of the stars.}
\end{figure}

\begin{figure}[p]
\includegraphics[width=12.7cm]{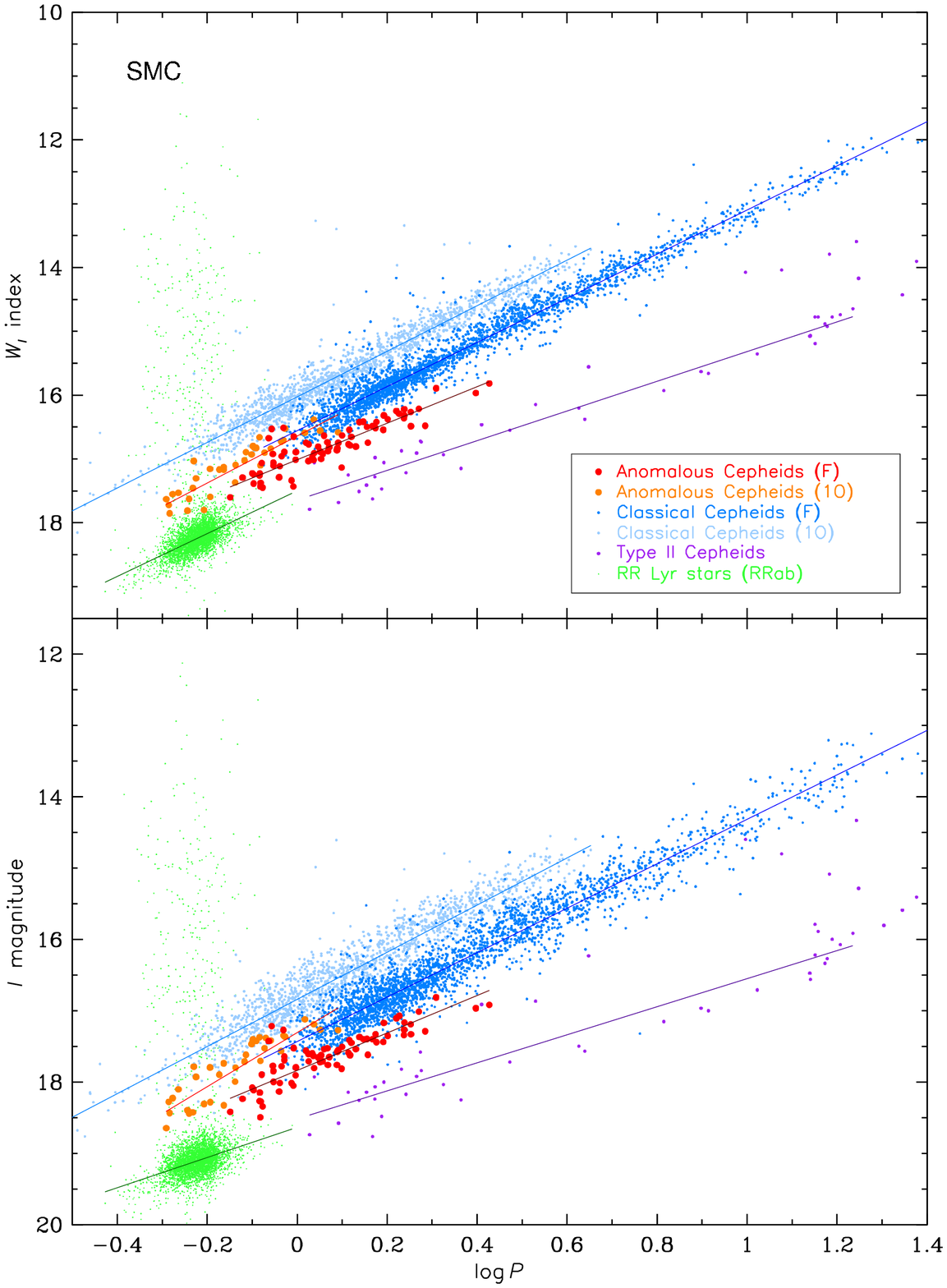}
\FigCap{Period--luminosity diagrams for Cepheids and RR Lyr stars in the
SMC. Symbols are the same as in Fig.~5.}
\end{figure}

\renewcommand{\TableFont}{\footnotesize}
\MakeTable{l@{\hspace{3pt}}
c@{\hspace{6pt}}
c@{\hspace{8pt}}
c@{\hspace{8pt}}
c@{\hspace{8pt}}
c@{\hspace{8pt}}
c@{\hspace{3pt}}}
{12.5cm}{Period--luminosity relations for pulsating stars in the Magellanic Clouds.}
{\hline
\multicolumn{1}{c}{Type of pulsators}
& Mode
& Galaxy
& $a$
& $\sigma_{a}$
& $b$
& $\sigma_{b}$ \\
\hline
\multicolumn{7}{c}{$W_I=a\log{P}+b$} \\
Anomalous Cepheids & F  & LMC & $-$3.05 & 0.11 & 16.59 & 0.02 \\
Anomalous Cepheids & 1O & LMC & $-$3.38 & 0.19 & 16.03 & 0.04 \\
Classical Cepheids & F  & LMC & $-$3.32 & 0.01 & 15.89 & 0.01 \\
Classical Cepheids & 1O & LMC & $-$3.43 & 0.01 & 15.39 & 0.01 \\
Type II Cepheids   & F  & LMC & $-$2.53 & 0.02 & 17.35 & 0.02 \\
RR Lyr stars       & F  & LMC & $-$3.06 & 0.02 & 17.15 & 0.01 \\
Anomalous Cepheids & F  & SMC & $-$2.85 & 0.15 & 17.01 & 0.03 \\
Anomalous Cepheids & 1O & SMC & $-$3.69 & 0.28 & 16.64 & 0.05 \\
Classical Cepheids & F  & SMC & $-$3.46 & 0.02 & 16.56 & 0.01 \\
Classical Cepheids & 1O & SMC & $-$3.57 & 0.02 & 16.03 & 0.01 \\
Type II Cepheids   & F  & SMC & $-$2.32 & 0.08 & 17.64 & 0.06 \\
RR Lyr stars       & F  & SMC & $-$3.35 & 0.06 & 17.50 & 0.02 \\
\hline
\multicolumn{7}{c}{$I=a\log{P}+b$} \\
Anomalous Cepheids & F  & LMC & $-$2.94 & 0.15 & 17.52 & 0.02 \\
Anomalous Cepheids & 1O & LMC & $-$3.31 & 0.22 & 16.83 & 0.05 \\
Classical Cepheids & F  & LMC & $-$2.91 & 0.02 & 16.82 & 0.01 \\
Classical Cepheids & 1O & LMC & $-$3.30 & 0.02 & 16.38 & 0.01 \\
Type II Cepheids   & F  & LMC & $-$2.00 & 0.04 & 18.20 & 0.03 \\
RR Lyr stars       & F  & LMC & $-$1.89 & 0.02 & 18.33 & 0.01 \\
Anomalous Cepheids & F  & SMC & $-$2.63 & 0.15 & 17.83 & 0.03 \\
Anomalous Cepheids & 1O & SMC & $-$3.78 & 0.34 & 17.31 & 0.06 \\
Classical Cepheids & F  & SMC & $-$3.12 & 0.02 & 17.43 & 0.01 \\
Classical Cepheids & 1O & SMC & $-$3.31 & 0.03 & 16.84 & 0.01 \\
Type II Cepheids   & F  & SMC & $-$1.96 & 0.12 & 18.51 & 0.09 \\
RR Lyr stars       & F  & SMC & $-$2.12 & 0.06 & 18.63 & 0.02 \\
\hline}

Figs.~5 and 6 show period \vs Wesenheit index (upper panels) and period \vs
{\it I}-band magnitude for Cepheids and RR Lyr stars in the LMC and SMC,
respectively. Wesenheit index is an extinction insensitive quantity,
defined as $W_I=I-1.55(V-I)$. As can be seen, ACs pulsating in the
fundamental and first-overtone modes follow the PL relations located below
their classical counterparts. In the LMC, anomalous and classical Cepheids
are well separated in the PL planes, while in the SMC both types of
pulsators seem to partly overlap, which was a source of difficulties in the
identifications of ACs in this environment.

The overlap of anomalous and classical Cepheids in the PL diagrams may be
partly explained by the large line-of-sight depth of the SMC, which
broadens the PL relations. However, the mean PL relations of ACs and
classical Cepheids seem to have inconsistent zero points in both
galaxies. We fitted the linear PL relations for ACs, classical Cepheids,
type II Cepheids (BL Her and W Vir stars) and RRab stars using a standard
least square method with $3\sigma$ clipping. Slopes and zero points of the
fits are summarized in Table~1. The fits are also shown in Figs.~5
and~6. We may compare mean luminosities of various types of Cepheids and RR
Lyr stars at the pulsation period of $P=1$~d ($\log{P}=0$), because all
studied types of pulsators have representatives with periods around this
value. In the LMC, fundamental-mode ACs are by $0.70\pm0.02$~mag fainter
than fundamental-mode classical Cepheids, while in the SMC this difference
is only $0.45\pm0.03$~mag (in the period--$W_I$ plane, but similar values
can be found in the PL diagrams in the {\it I} band). Similar discrepancy
is also visible when we compare classical Cepheids and RR Lyr stars or type
II Cepheids, which suggests that ACs belong to the old population. The
effect of different zero points of the PL relations in the LMC and SMC is
not visible for the first-overtone Cepheids, but it cannot be excluded that
the overtone ACs in the SMC are affected by the selection effect -- the
brightest ones still remain on the list of classical Cepheids.

Assuming that young and old stars in the SMC are on average at the same
distance from us, the proximity of anomalous and classical Cepheids in the
SMC has consequences for the calibration of Cepheids as distance
indicators. Either classical Cepheids in the SMC are fainter than in the
LMC, or ACs in the SMC are brighter than in the LMC. It is also possible
than both conjectures are true.

Note that the reclassification of several dozen short-period classical
Cepheids as ACs changed the appearance of the PL diagram for Cepheids in
the SMC. The break in the slope observed for the fundamental-mode classical
Cepheids at a period of about 2.5~d ($\log{P}=0.4$, Bauer \etal 1999, Udalski
\etal 1999) is not so distinct when ACs were separated, although it is
still visible. Whatever the cause of this non-linearity (see Bauer \etal
1999 for the discussion of possible explanations of such behavior of
Cepheids), this effect additionally hampers the detection of ACs in the
SMC, since short-period classical Cepheids approach ACs in the PL plane.

\begin{figure}[t]
\begin{center}
\includegraphics[width=12.7cm]{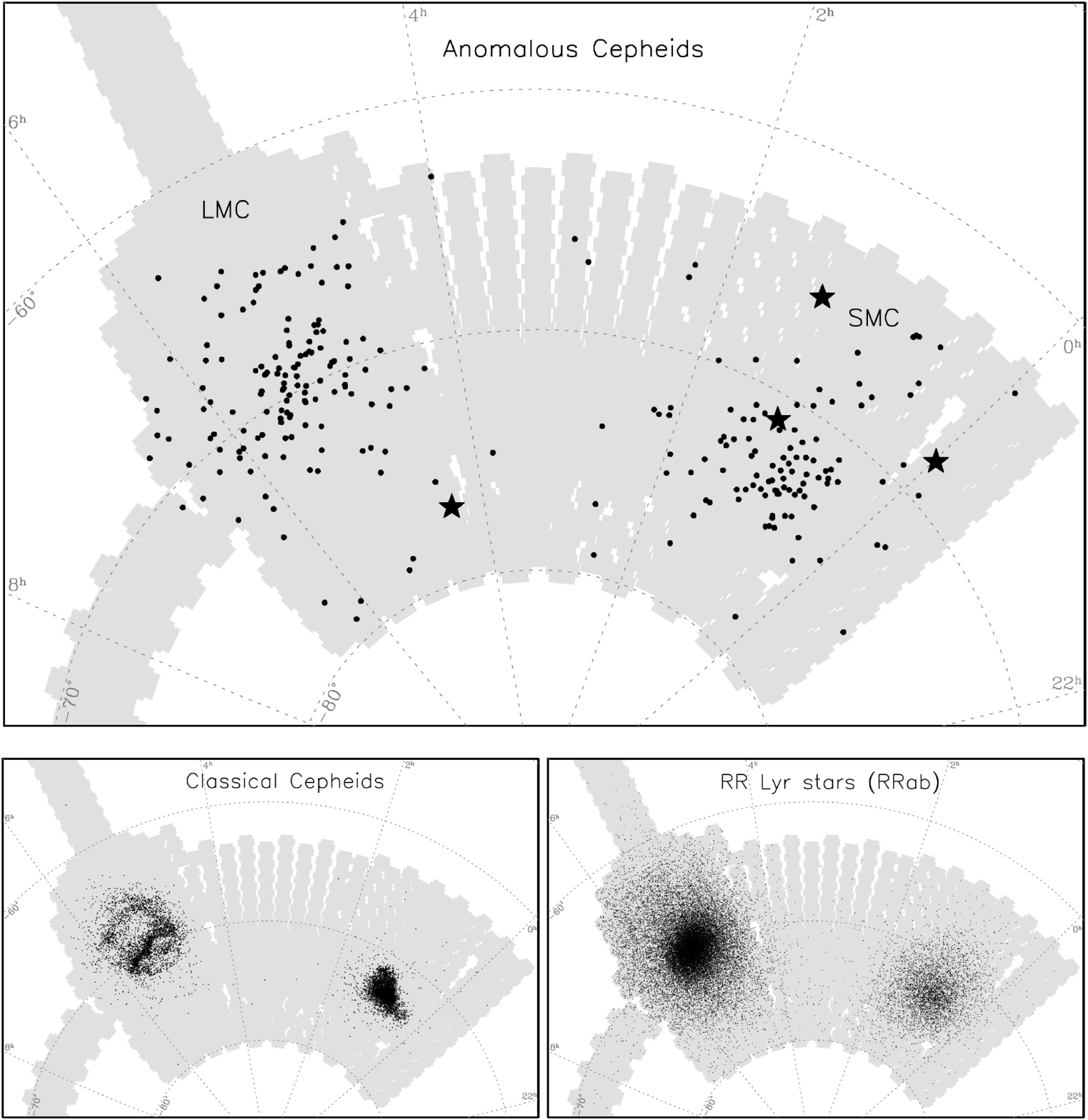}
\end{center}
\vspace*{-3mm}
\FigCap{Spatial distribution of ACs ({\it upper panel}), classical Cepheids
({\it lower left panel}), and RRab stars ({\it lower right panel}) in the
Magellanic System. Star symbols show Galactic ACs in the foreground of the
Magellanic Clouds. The grey area shows the sky coverage of the OGLE fields.}
\end{figure}

\Subsection{Spatial Distribution of Anomalous Cepheids in the Magellanic Clouds}

Study of spatial distribution of various stars is a powerful tool to
discriminate between old, intermediate and young stellar
populations. Fiorentino and Monelli (2012) suggested that a survey for ACs
in the outer regions of the LMC may solve the problem of their
origin. OGLE-IV covers outskirts of the Magellanic Clouds, including the
Magellanic Bridge -- a region located between them.

In the upper panel of Fig.~7 we present the positions of our sample of 250
ACs in the Magellanic System. In the lower panels of Fig.~7 we plot the
positions of classical Cepheids (tracers of the young stellar population)
and fundamental-mode RR Lyr stars (representatives of the old population)
found in the same OGLE fields toward the Magellanic Clouds. Our collection
of classical Cepheids in the Clouds contains in total 9513 objects. Our
current sample of RRab stars (fundamental-mode RR Lyr stars) detected in
the OGLE fields toward the Magellanic Clouds consists of $32\;483$
objects. In this study we used RRab stars only, since their light curves
have very characteristic shapes and we expect the largest completeness and
the least contamination in this sample. To reject RR Lyr variables
belonging to the Milky Way halo, we removed objects more than 0.6~mag
brighter or fainter than the mean fit to the PL relation. In our last cut
we removed RR Lyr stars that likely belong to globular clusters in both
Clouds. Our final sample counts $30\;331$ RRab stars in the LMC and SMC.

The difference between spatial distributions of the young and old stellar
population is clearly visible in the lower panels of Fig.~7. Against this
background, the distribution of ACs is ambiguous. In the LMC, ACs seem to
trace the bar and spiral arms of this galaxy, like classical Cepheids,
although there is an excess of objects far from the LMC center, in
particular in the southern direction. In the SMC, ACs seem to follow very
broad spatial distribution, similarly to the RR Lyr stars. Several ACs can
be found even in the Magellanic Bridge, at halfway between Clouds.

\begin{figure}[t]
\begin{center}
\includegraphics[width=12.7cm]{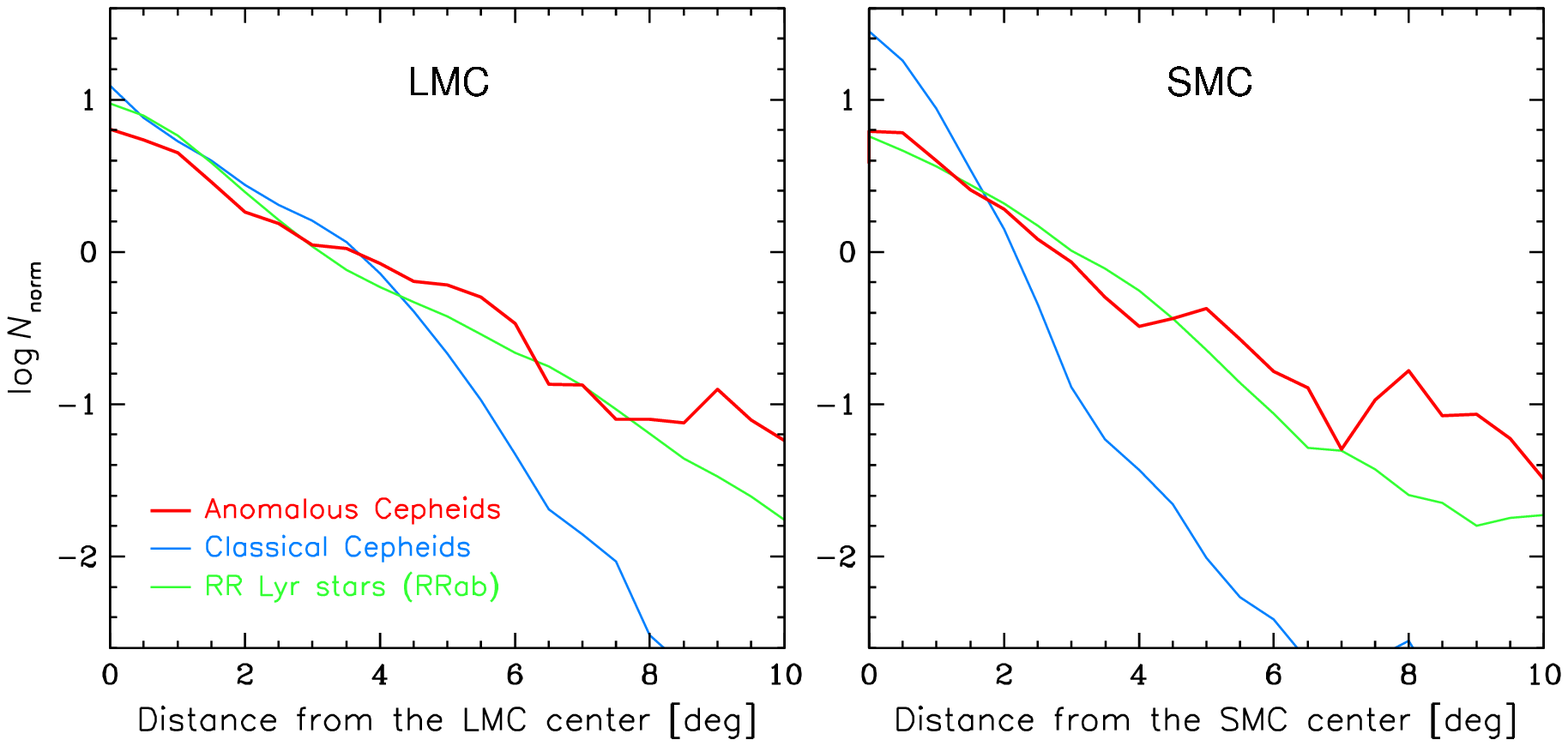}
\end{center}
\vspace*{-3mm}
\FigCap{One-dimensional distributions of ACs (red lines), classical
Cepheids (blue lines), and RRab stars (green lines) plotted against
angular distance from the LMC ({\it left panel}) and SMC center ({\it right
panel}). The number of classical Cepheids and RR Lyr stars were normalized
to the number of ACs within 10~degrees from the galactic centers.}
\end{figure}

We derived one-dimensional density distributions of ACs, classical Cepheids
and RRab stars in both Clouds. We counted the number of pulsators in
one-degree-wide circular rings around the centers of the LMC
($\alpha_{LMC}=80.8942^{\circ}$, $\delta_{LMC}=-69.7561^{\circ}$) and SMC
($\alpha_{SMC}=13.1866^{\circ}$, $\delta_{SMC}=-72.8286^{\circ}$) and
rescaled the numbers of classical Cepheids and RR Lyr stars to the number
of ACs within 10~degrees from the galactic centers. Taking into account the
incomplete sky coverage of the OGLE fields, we derived the number of stars
per square degree in each ring. These numbers in the logarithmic scale are
plotted in Fig.~8 against the projected angular distances from the centers
of the LMC and SMC. 

In both galaxies, the radial distribution of ACs resembles that of RR Lyr
stars. It is particularly evident in the SMC, where classical Cepheids are
much more concentrated in the galactic center than the old stellar
population. In both Clouds, there is an excess of ACs with respect to RR
Lyr stars for angular distances larger than 7~degrees from the
centers. This suggests that at least part of the ACs in the Magellanic
Clouds are very old stars, which implies their binary origin. On the other
hand, the traces of the bar and spiral arm visible in the distribution of
ACs in the LMC may indicate much younger population. The most likely
explanation is that ACs in the Magellanic Clouds represent a mixture of two
formation channels: the evolution of coalescent binaries and the evolution
of single, intermediate-age, metal-poor stars.

\Subsection{Galactic Anomalous Cepheids in the Foreground of the Magellanic Clouds}

\begin{figure}[t]
\begin{center}
\includegraphics[width=12.5cm]{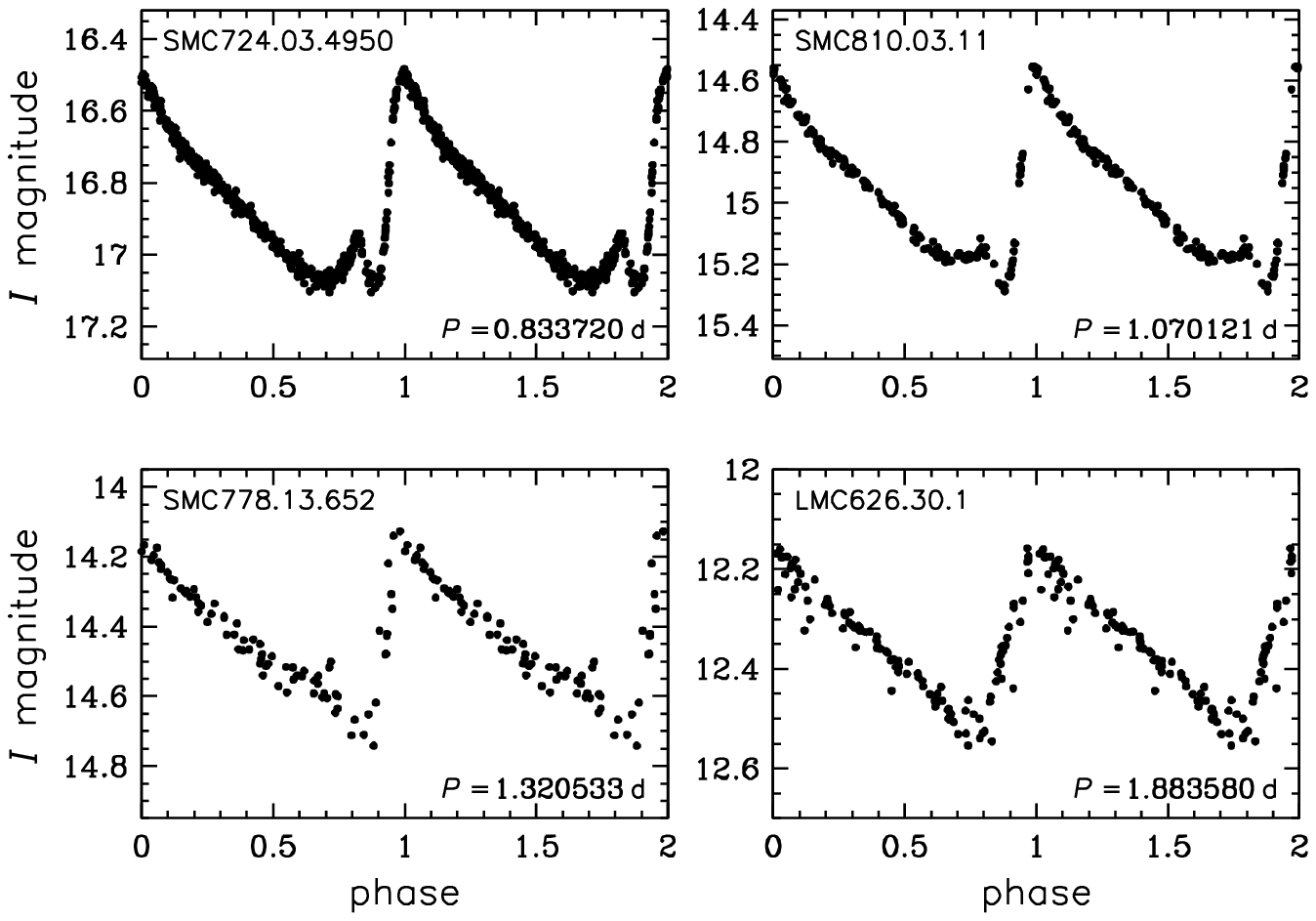}
\end{center}
\vspace*{-3mm}
\FigCap{{\it I}-band light curves of four Galactic ACs in the foreground of
the Magellanic Clouds.}
\end{figure}

Four pulsating stars found in the OGLE fields toward the Magellanic Clouds
are much brighter than expected for fundamental-mode Cepheids from this
galaxies (Fig.~5), but have light curves typical for ACs (Fig.~9). The
Fourier coefficients $\phi_{21}$ and $\phi_{31}$ of the decomposed light
curves place them among fundamental-mode ACs in the Magellanic Clouds
(Fig.~1). Given that the Magellanic Clouds are far from the plane of the
Milky Way (so we do not expect to find Galactic classical Cepheids in
these regions), we can safely assume that these four objects are ACs
belonging to our Galaxy. Until now, the only known AC in Galactic field was
a first-overtone pulsator XZ Ceti (Szabados \etal 2007). The OGLE variables
are the first known fundamental-mode ACs found in the field of the Milky
Way. Their basic observational parameters are summarized in Table~2.
Individual distances to these stars were derived from the $\log{P}$--$W_I$
relation of the LMC ACs, assuming the distance to the LMC of 49.97~kpc
(Pietrzyñski \etal 2013).

\renewcommand{\TableFont}{\footnotesize}
\MakeTable{l@{\hspace{3pt}}
c@{\hspace{6pt}}
c@{\hspace{8pt}}
c@{\hspace{8pt}}
c@{\hspace{8pt}}
c@{\hspace{8pt}}
c@{\hspace{3pt}}
c@{\hspace{3pt}}}
{12.5cm}{Galactic anomalous Cepheids in the foreground of the Magellanic Clouds.}
{\hline
\multicolumn{1}{c}{Identifier}
& Pulsation
& $P$
& $\langle{I}\rangle$
& $\langle{V}\rangle$
& R.A.
& Dec.
& Distance \\
& mode
& [d]
& [mag]
& [mag]
& [J2000.0]
& [J2000.0]
& [kpc] \\
\hline
SMC724.03.4950 & F & 0.8337199 & 16.850 & 17.373 & 01:10:57.51 & $-$71:01:57.6 & 34.7 \\
SMC810.03.11   & F & 1.0701211 & 14.955 & 15.387 & 01:20:14.74 & $-$65:42:38.3 & 18.0 \\
SMC778.13.652  & F & 1.3205328 & 14.443 & 14.939 & 23:59:14.51 & $-$68:13:56.6 & 15.3 \\
LMC626.30.1    & F & 1.8835803 & 12.344 & 13.015 & 04:23:51.47 & $-$76:54:42.8 & ~~6.5 \\
\hline}

Recently, Sipahi \etal (2013ab) measured masses of two pulsating stars that
are components of eclipsing binary systems. Their relatively small masses
-- 1.64 and 1.46 $M_\odot$ -- led the authors to the conclusion that both
pulsators are anomalous Cepheids. However, their long pulsation periods --
4.15 and 4.22~d, respectively -- suggest that both objects belong to a
different class of pulsating stars: the so called peculiar W~Vir stars
discovered by Soszyñski \etal (2008) in the LMC. Indeed, OGLE light curves
show that a large fraction (at least 30\%) of peculiar W~Vir stars are
members of binary systems.

\Section{Conclusions}

We presented a collection of 250 ACs in the Magellanic Clouds, which
probably exceeds the total number of ACs known in all other galaxies. Such
a big sample of these rare pulsating stars is sufficient to perform various
statistical tests concerning their evolution, PL relations, spatial
distributions, features of the light curves, metallicities, etc. In the
LMC, ACs and classical Cepheids clearly constitute two separate classes of
variable stars, while in the SMC both groups of Cepheids partly overlap in
the PL diagram and exhibit much more similar light curves, which is
associated with a lower metallicity of classical Cepheids in the SMC than
in the LMC. This similarity could confirm the suggestion of Caputo \etal
(2004) that ACs are natural extension of classical Cepheids to lower metal
contents and smaller masses, so ACs do not constitute in fact a separate
class of variable stars. However, we demonstrated that ACs and classical
Cepheids in the SMC follow completely different spatial distributions
which, in turn, indicates different evolutionary histories of both groups
of pulsators.

The discovery of four fundamental-mode ACs in the foreground of the
Magellanic Clouds proves that ACs are quite numerous in the Galactic halo,
but until now there was no good method to distinguish them from other types
of Cepheids. We showed that Fourier parameters $\phi_{21}$ and $\phi_{31}$
are useful discriminants of ACs and classical Cepheids.

\Acknow{We are grateful to Z.~Ko³aczkowski and A.~Schwarzenberg-Czerny for
providing software which enabled us to prepare this study.

This work has been supported by the Polish Ministry of Science and Higher
Education through the program ``Ideas Plus'' award No. IdP2012 000162. The
OGLE project has received funding from the Polish National Science Centre
grant MAESTRO no. 2014/14/A/ST9/00121 to AU.}

\end{document}